\pdfoutput=1

\documentclass[11pt]{article}

\usepackage[final]{acl}

\usepackage{times}
\usepackage{latexsym}
\usepackage[T1]{fontenc}
\usepackage[utf8]{inputenc}
\usepackage{microtype}
\usepackage{inconsolata}

\usepackage{color,xcolor}
\usepackage{epsfig}
\usepackage{graphicx}
\usepackage{lipsum}
\usepackage{array}
\usepackage{booktabs}
\usepackage{colortbl}
\usepackage{multirow}
\usepackage{float}
\usepackage{caption}
\usepackage{adjustbox}
\usepackage{subcaption}
\usepackage{footnote}
\makesavenoteenv{tabular}
\makesavenoteenv{table}
\usepackage{makecell}
\usepackage{tablefootnote}

\usepackage{amsmath,amsfonts,amssymb,bm}
\usepackage[super]{nth}

\usepackage{changepage}
\usepackage{extramarks}
\usepackage{lastpage}
\usepackage{setspace}
\usepackage{soul}
\usepackage{xspace}

\usepackage{url}
\usepackage{hyperref}
\hypersetup{colorlinks=True, urlcolor=black}

\usepackage[ruled,vlined]{algorithm2e}
\usepackage{enumitem}
\usepackage{verbatim}
\usepackage{pifont}
\usepackage[acronyms]{glossaries}
\glsdisablehyper
\def\ours{IST-LM}

\title{Interleaved Speech-Text Language Models for Simple Streaming Text-to-Speech Synthesis}

\author{
\textbf{Yifan Yang\textsuperscript{1}\thanks{*Work done during an internship at Microsoft.}},
\textbf{Shujie Liu\textsuperscript{2}},
\textbf{Jinyu Li\textsuperscript{2}},
\textbf{Hui Wang\textsuperscript{2}},
\textbf{Lingwei Meng\textsuperscript{2}},
\textbf{Haiyang Sun\textsuperscript{2}} \\
\textbf{Yuzhe Liang\textsuperscript{1}},
\textbf{Ziyang Ma\textsuperscript{1}},
\textbf{Yuxuan Hu\textsuperscript{2}},
\textbf{Rui Zhao\textsuperscript{2}},
\textbf{Jianwei Yu\textsuperscript{2}},
\textbf{Yan Lu\textsuperscript{2}},
\textbf{Xie Chen\textsuperscript{1}}
\\
\textsuperscript{1}Shanghai Jiao Tong University \\
\textsuperscript{2}Microsoft Corporation
}

\begin{document}

\maketitle

\begin{abstract}
This paper introduces Interleaved Speech-Text Language Model (IST-LM) for zero-shot streaming Text-to-Speech (TTS). Unlike many previous approaches, IST-LM is directly trained on interleaved sequences of text and speech tokens with a fixed ratio, eliminating the need for additional efforts like forced alignment or complex designs. The ratio of text chunk size to speech chunk size is crucial for the performance of IST-LM. To explore this, we conducted a comprehensive series of statistical analyses on the training data and performed correlation analysis with the final performance, uncovering several key factors: 1) the distance between speech tokens and their corresponding text tokens, 2) the number of future text tokens accessible to each speech token, and 3) the frequency of speech tokens precedes their corresponding text tokens. Experimental results demonstrate how to achieve an optimal streaming TTS system with a limited performance gap compared to its non-streaming counterpart. IST-LM is conceptually simple and empirically powerful, enabling streaming TTS with minimal overhead while largely preserving performance, and offering broad potential for integration with real-time text streams from large language models.
\end{abstract}
\section{Introduction}
Text-to-speech (TTS) synthesis, which aims to generate high-fidelity speech from text, has made remarkable progress, driven by advancements in generative modeling~\cite{tacotron2, transformertts, vits, fastspeech2, difftts, valle}, as well as the growing availability of computational power and data~\cite{wenetspeech4tts, libriheavy, emilia, gigaspeech, gigaspeech2}. Consequently, modern TTS systems exhibit human-level parity in terms of naturalness and intelligibility, for both predefined  speakers~\cite{naturalspeech} and zero-shot scenarios~\cite{valle2}.

While existing zero-shot TTS systems~\cite{valle2, melle, cosyvoice, maskgct, e2tts, f5tts} demonstrate promising performance in synthesizing speech for unseen speakers, they are typically trained in an offline mode and process the entire input text before generating speech. As a result, these systems suffer from high latency and prohibitive computational costs when handling very long texts.
To mitigate these challenges, existing zero-shot streaming TTS systems~\cite{livespeech, livespeech2} break long text inputs into smaller chunks and synthesize speech for each chunk separately. However, this leads to inconsistencies across different chunks. There remains substantial room for improving streaming TTS.

A more intuitive but less explored solution to this challenge involves interleaving text and speech tokens at a fixed ratio. This strategy leverages the in-context learning (ICL) capabilities of language models (LMs) to ensure consistent timbre and prosody across speech segments while aligning naturally with the steady output rate of large language models (LLMs).

With this perspective in mind, this paper introduces \textbf{I}nterleaved \textbf{S}peech-\textbf{T}ext \textbf{L}anguage \textbf{M}odel (\ours{}) for zero-shot streaming TTS, a novel paradigm that directly trains an LM on interleaved sequences of text and speech tokens with a fixed ratio. This eliminates the need for additional efforts such as forced alignment to prepare training data and complex system designs.
To investigate the key factors involved in the interleaving design, specifically chunk-internal size and chunk-mutual ratio, we propose four sets of word-level, position-aware statistical measures, and perform statistical analyses on the entire training dataset. By correlating these measures with the final model performance, we uncover several key insights:
\begin{itemize}[leftmargin=*, itemsep=0pt, topsep=0pt]
\item The ratio of text chunk size to speech chunk size directly affects 1) the distance between speech tokens and their corresponding text tokens, 2) the number of future text tokens accessible to each speech token, and 3) the frequency of speech tokens preceding their corresponding text tokens.
\item The mean distance between speech tokens and their corresponding text tokens reflects a trade-off: shorter distances impose stronger constraints on speech synthesis while limiting the available contextual information as fewer upcoming text tokens are accessible to the current speech token, further impacting model performance.
\item The variance in the distances between speech tokens and their corresponding text tokens indicates the modeling difficulty of the LM. When the chunk-mutual ratio is fixed, the variance changes very little.
\item The frequency of speech tokens preceding their corresponding text tokens is highest at the start of the interleaved sequence, increasing modeling difficulty during training due to the lack of context from text tokens. However, this typically does not affect inference with speech prompts.
\end{itemize}
Experiments conducted on LibriTTS, using the LibriSpeech test-clean set for zero-shot TTS evaluation, demonstrate that \ours{} with a $1\colon3$ ratio increases the word error rate (WER) by only 8\% relatively compared to the non-streaming counterpart while maintaining comparable speaker similarity and overall perceived quality.
\ours{} is conceptually simple and empirically powerful, presenting a promising solution for streaming TTS. We hope that our streaming TTS model and the insights derived from our analysis will contribute to the advancement of the voice interaction field.
\section{Related Work}
\subsection{Speech Language Models}
The advent of LLMs has spurred the integration of multiple modalities by converting them into continuous or discrete tokens for joint training, which has emerged as a promising approach. Previous studies have explored the joint modeling of speech and text for various applications, including automatic speech recognition (ASR)~\cite{slamasr, seedasr}, text-to-speech synthesis (TTS)~\cite{cosyvoice, seedtts}, and voice dialog systems~\cite{speechgpt, zeng2024scaling}.
In these studies, some approaches treat text and speech tokens separately, with text tokens guiding speech tokens~\cite{cosyvoice, seedtts}, or speech tokens guiding text tokens~\cite{slamasr, speechllama, seedasr}. Other approaches interleave text and speech tokens. SpiritLM~\cite{spiritlm} randomly replaces paired speech and text token spans to enhance modality switching during generation, while ELLAV~\cite{ellav} interleaves phonemes and their corresponding speech tokens to enforce the constraint of text-to-speech synthesis. However, these two methods depend heavily on forced alignment, which introduces additional computational overhead and poses challenges for scalability. GLM-4-Voice~\cite{glm4voice} is pre-trained on interleaved sequences of text and corresponding synthesized speech data, bypassing forced alignment, yet the speech and text chunks remain paired during training.
OmniFlatten~\cite{omniflatten} is trained on interleaved dialogue sequences of text and speech chunks with fixed sizes, where the text chunk size is 2 and the speech chunk size is 10. However, these chunk sizes are large and empirically chosen, with the ratio selected solely to prevent the output text from excessively preceding the speech content, lacking a deeper exploration or analysis of alternative ratios.
The investigation of interleaving speech and text tokens at a fixed ratio remains limited.

\subsection{Zero-Shot TTS}
Zero-shot TTS systems enable speech synthesis for unseen speakers by capturing the timbre, prosody, and style from merely several seconds of speech prompts. Early approaches primarily focus on speaker adaptation~\cite{NeuralVoiceCloning, SampleEfficientTTS, adaspeech} and speaker encoding~\cite{TransferLearningTTS}, often requiring model fine-tuning, feature engineering, or complex structural designs.
As language modeling rapidly advances, the performance of zero-shot TTS systems has greatly improved, achieving human-level quality in naturalness and intelligibility~\cite{valle2}.

Recent research in zero-shot TTS can be broadly classified into two categories: some use speech prompts~\cite{valle, valle2, vallt, melle, felle, maskgct, e2tts} or speaker vectors~\cite{basetts} for in-context learning (ICL), and others disentangle speaker information from speech signals~\cite{naturalspeech3}.
More recent works~\cite{cosyvoice, cosyvoice2, palle} combine speaker disentanglement and ICL to achieve better performance.

\subsection{Streaming TTS}
Streaming TTS systems incrementally convert incoming text into a speech stream, aiming to reduce perceived latency, particularly for long inputs, by enabling audio playback before the entire text is processed. With the advent of LLMs, streaming TTS has been adapted for real-time voice synthesis from LLM outputs, improving the naturalness of voice interactions and enhancing user experience.

Existing streaming TTS systems can be broadly divided into chunk-level and frame-level generation methods. Traditional chunk-level systems~\cite{llm2speech, livespeech} segment the long text into chunks based on punctuation or word boundaries, and synthesize speech for each chunk separately, leading to inconsistencies and unnatural transitions across chunks. Subsequent work~\cite{livespeech2} adopts sliding window and context pruning to alleviate these issues. Nevertheless, these methods heavily rely on complex rule-based segmentation and engineering optimization. Frame-level systems leverage the inherently streaming nature of neural Transducers~\cite{RNN-T}. Early work like Speech-T~\cite{speecht} focuses on single-speaker synthesis without zero-shot capabilities. Several recent approaches~\cite{vallt, ttstransducer, TransducerAndSpeak, TokenTransducerTTS} incorporate zero-shot capabilities but are primarily designed for non-streaming scenarios.

Given that LLMs generate text at a constant rate, there is considerable potential for developing more efficient streaming TTS systems without intricate engineering efforts. This naturally raises the question: \textbf{\textit{can speech be synthesized in parallel with LLM-generated text at a fixed ratio?}} In this work, we explore the feasibility of interleaving text and speech tokens with a fixed ratio, demonstrating its potential in voice dialogue systems.

\begin{figure*}[t!]
  \centering
  \includegraphics[width=0.95\linewidth]{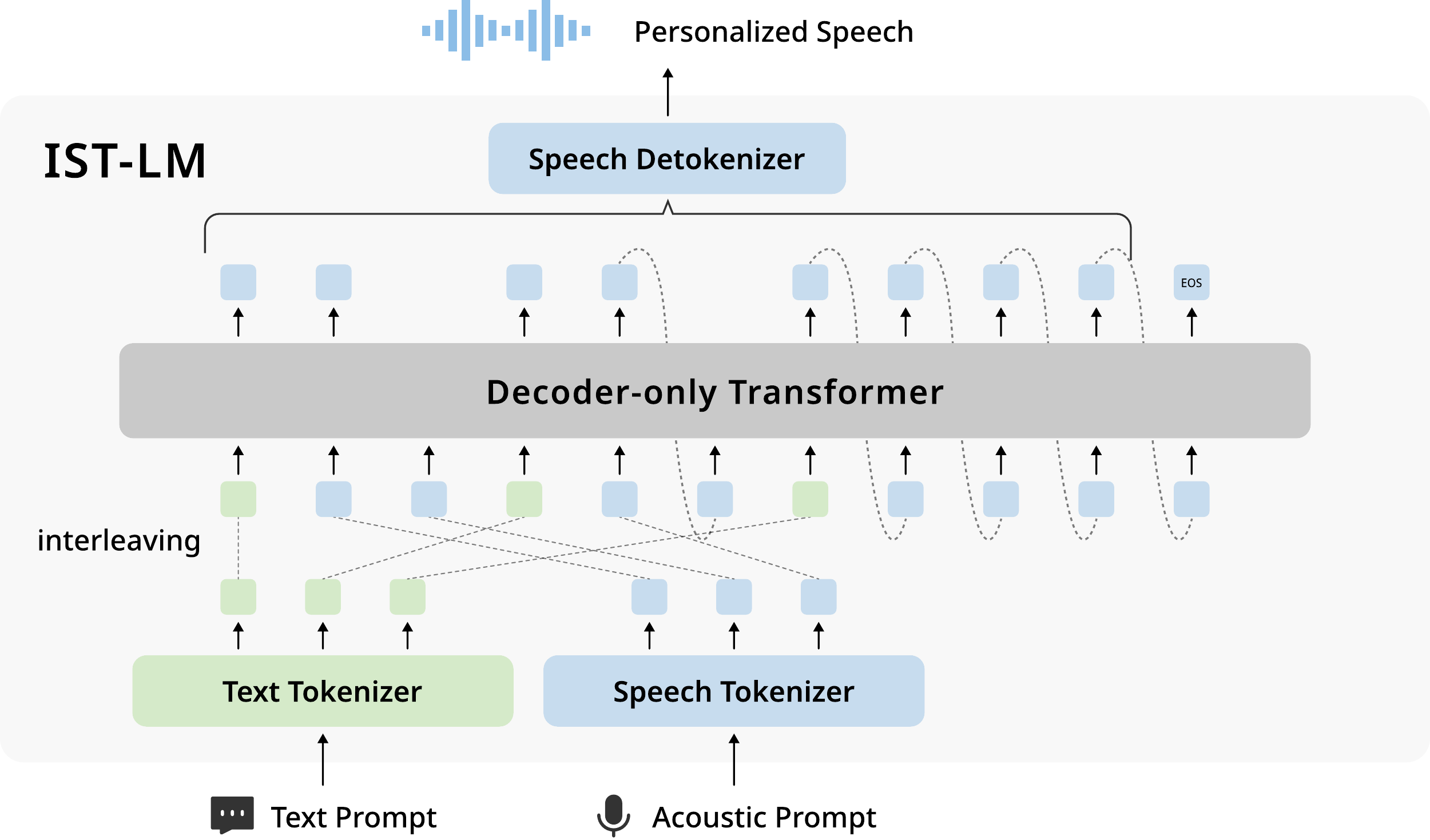}
  \caption{An overview of the proposed \ours{} model, comprising (1) a BPE-based text tokenizer, (2) a supervised speech tokenizer, (3) a decoder-only LM modeling interleaved sequence of speech and text tokens with a fixed ratio ($1 \colon 2$ is used for illustration in the figure) as input, and (4) a conditional flow matching decoder with a vocoder.}
  \label{fig:istlm}
\end{figure*}

\section{Problem Formulation: Regarding Streaming TTS as Interleaved Speech-Text Language Modeling}
Streaming TTS systems are required to continuously synthesize speech segments from an incoming text stream, generating speech outputs in real-time scenarios.
In this paper, we regard zero-shot streaming TTS as an interleaved speech-text language modeling task, treating streaming speech synthesis as a joint sequential modeling problem.

\noindent\textbf{Formulation}\quad
Consider a speech sample $\mathbf{y}$ and its corresponding transcription $\mathbf{x}$. The transcription $\mathbf{x}$ is converted into subword units using Byte Pair Encoding (BPE)~\cite{bpe}, resulting in BPE token sequence $\bm{x} = [x_0, x_1, \ldots, x_{S-1}]$, where $S$ is the length of the tokenized sequence.
A pre-trained speech tokenizer is used to encode the speech sample into speech tokens, denoted as $\bm{y}  = [y_0, y_1, \ldots, y_{T-1}] = \text{Encode}_\text{spch}(\mathbf{y})$, where $\bm{y}$ represents the speech token sequence of downsampled length $T$.
After quantization, a pre-trained speech detokenizer along with a vocoder can reconstruct the waveform, denoted as $\text{Decode}_\text{spch}(\bm{y}) \approx \hat{\bm{y}}$.

We train a neural LM on the interleaved sequence of BPE tokens $\bm{x}$ and speech tokens $\bm{y}$ with a predefined fixed ratio of $n \colon m$. The interleaved sequence $\bm{l}$ is constructed as follows:
\begin{equation}
\bm{l} = [x_{0:n-1}, y_{0:m-1}, x_{n:2n-1}, y_{m:2m-1}, \ldots], 
\end{equation}
where the BPE tokens and speech tokens are alternated in blocks of size $n$ and $m$, respectively. Once the BPE tokens are consumed, the remaining speech tokens are appended to the end of the sequence.
The LM is optimized to predict this interleaved sequence $\bm{l}$ using cross-entropy loss. Specifically, at each timestep $t$, the LM is expected to predict the next speech token $y_t$ conditioned on the previously generated sequence $\bm{l}_{<t}$. The optimization objective is:
\begin{equation}
\arg\max_{\theta} p(\bm{l}_t \mid \bm{l}_{<t}; \theta),
\end{equation}
where $\bm{l}_{<t}$ represents the sequence $[l_0, l_1, \ldots, l_{t-1}]$, and $\theta$ denotes the parameters of the LM. Notably, only losses for speech tokens are computed.

During inference, given the BPE tokens $\bm{x}$ of the text to be synthesized, the speech tokens $\Tilde{\bm{y}}$ from the speech prompt, and the BPE tokens $\Tilde{\bm{x}}$ of the corresponding text prompt, the LM generates the target speech tokens $\bm{y}$ in a streaming manner while preserving the speaker characteristics of the speech prompt. The BPE tokens $\bm{x}$ and $\Tilde{\bm{x}}$ are concatenated and treated as a unified sequence, which is then segmented into chunks of size $n$. For each chunk of $n$ BPE tokens, the model generates $m$ speech tokens, repeating this process until either the $\texttt{<EOS>}$ token is produced or all BPE tokens are consumed. In the latter case, the model continues to generate the remaining speech tokens sequentially until the $\texttt{<EOS>}$ token is emitted.


\section{\ours{}}
\subsection{Architecture}
The overall architecture of \ours{} is illustrated in Fig.~\ref{fig:istlm}. \ours{} comprises the following main components: a BPE-based text tokenizer that converts raw text into sub-word tokens; a speech tokenizer that encodes speech samples into discrete speech tokens; a decoder-only LM that models interleaved sequences of speech and text tokens; a speech detokenizer with a built-in vocoder that synthesizes waveform from the speech tokens.

\subsection{Speech Tokenization and Detokenization}
For speech tokenization, we utilize the pre-trained S3Tokenizer from CosyVoice~\cite{cosyvoice} to extract discrete semantic speech tokens from the waveform at a token rate of 50 Hz. This model is a fine-tuned version of the SenseVoice-Large~\cite{funaudiollm} ASR model, which is trained on a large multilingual speech dataset, providing robust speech understanding capabilities. By leveraging ASR loss during training, the S3Tokenizer can extract semantic information while disregarding irrelevant noise and speaker information. This enables the S3Tokenizer to implicitly denoise and disentangle speakers~\cite{touchtts}.

For speech detokenization, we adopt the pre-trained optimal-transport conditional flow matching model (OT-CFM) from CosyVoice~\cite{cosyvoice} to decode speech tokens into mel spectrograms, which are then transformed into the waveform using the pre-trained HiFi-GAN~\cite{hifi-gan} vocoder from CosyVoice~\cite{cosyvoice}.

\subsection{Interleaved Speech-Text Language Model}
We use a unidirectional Transformer decoder as the LM to autoregressively generate discrete speech tokens from the interleaved sequence of text and speech tokens with a fixed ratio. Input text tokens, appended with an $\texttt{<EOS>}$ token, are embedded via the text embedding layer, while speech tokens are projected into the semantic space of LM through the acoustic embedding layer.
By using distinct positional encodings for text and speech, the LM clearly distinguishes between the two modalities, leveraging multi-head attention and feed-forward layers to capture dependencies between semantic and acoustic information.
\begin{table*}[t]
\small
\centering
\caption{Objective performance comparison on continuation and cross-sentence zero-shot speech synthesis tasks.
\ours{}$_{n:m}$ represents streaming systems with a text chunk size of $n$ and a speech chunk size of $m$, while \ours{}$_{\infty:\infty}$ refers to non-streaming system.
\textbf{Bold} highlights the best result among \textbf{streaming systems}, while \underline{underlined} marks the second-best. $^*$Metrics not reported in the original papers are calculated using the checkpoints provided by their authors.
}

\renewcommand\tabcolsep{10pt}
\resizebox{0.95\linewidth}{!}{
\begin{tabular}{lccccc}
\toprule
\multirow{2}{*}{\textbf{System}} & \multicolumn{2}{c}{\textbf{Continuation}} & \multicolumn{3}{c}{\textbf{Cross-Sentence}} \\
\cmidrule(r){2-3} \cmidrule(r){4-6} & WER-H$\downarrow$ & SIM-o$\uparrow$ & WER-H$\downarrow$ & SIM-o$\uparrow$ & RTF$\downarrow$ \\
\midrule
Ground Truth                       & 2.15 & 0.905 & 2.15 & 0.779 & - \\
Ground Truth (EnCodec)             & 2.33 & 0.823 & 2.33 & 0.715 & - \\
Ground Truth (S3Tokenizer v1 50Hz) & 2.94 & 0.791 & 3.09 & 0.746 & - \\
\midrule
\textbf{Trained on Large-Scale Dataset} \\ 
VALL-E~\cite{valle}                & 3.80 & 0.773 & 5.90 & 0.633 & 0.73 \\
MaskGCT$^*$                        & -    & -     & 4.22 & 0.756 & 0.65 \\
E2 TTS (32 NFE)$^*$                & -    & -     & 2.92 & 0.756 & 0.68 \\
\midrule
\textbf{Trained on Small-Scale Dataset} \\
VALL-E                             & 4.47             & 0.730             & 8.64             & 0.531 & 0.73 \\
\ours{}$_{\infty:\infty}$          & 3.35             & 0.756             & 4.16             & 0.652 & 0.40 \\
\ours{}$_{1:2}$                    & 3.69             & 0.754             & 4.61             & 0.649 & 0.40 \\
\ours{}$_{1:3}$                    & \textbf{3.60}    & \underline{0.757} & \textbf{4.53}    & \textbf{0.653} & 0.40 \\
\ours{}$_{1:4}$                    & 5.73             & \underline{0.757} & 6.86             & 0.645 & 0.40 \\
\ours{}$_{3:6}$                    & 3.77             & \underline{0.757} & 5.26             & 0.650 & 0.40 \\
\ours{}$_{3:9}$                    & \underline{3.65} & \underline{0.757} & 4.75             & \underline{0.652} & 0.40 \\
\ours{}$_{3:12}$                   & 3.89             & \underline{0.757} & 5.20             & 0.649 & 0.40 \\
\ours{}$_{6:12}$                   & 3.76             & \textbf{0.758}    & 5.86             & 0.650 & 0.40 \\
\ours{}$_{6:18}$                   & 3.71             & 0.755             & 5.38             & 0.647 & 0.40 \\
\ours{}$_{6:24}$                   & 5.74             & 0.753             & 8.90             & 0.643 & 0.40 \\
\ours{}$_{12:24}$                  & 3.86             & \underline{0.757} & 5.96             & 0.646 & 0.40 \\
\ours{}$_{12:36}$                  & 3.70             & 0.754             & 5.58             & 0.649 & 0.40 \\
\ours{}$_{12:48}$                  & 3.80             & 0.756             & 5.19             & 0.646 & 0.40 \\
\bottomrule
\end{tabular}
}
\label{tab:main}
\end{table*}

\begin{table}[ht]
\small
\centering
\caption{
Predicted MOS comparison on cross-sentence zero-shot speech synthesis tasks.
}

\begin{tabular}{lc}
\toprule[1pt]
\textbf{System} & UTMOSv2$\uparrow$  \\
\midrule
Ground Truth & 3.22 \\
Ground Truth (S3Tokenizer v1 50Hz) & 3.30 \\
\midrule
\textbf{Trained on Large-Scale Dataset} \\
MaskGCT                   & 2.92 \\
E2 TTS (32 NFE)           & 2.82 \\
\midrule
\textbf{Trained on Small-Scale Dataset} \\ 
VALL-E                    & 2.12 \\
\ours{}$_{\infty:\infty}$ & \textbf{3.32} \\
\ours{}$_{1:3}$           & 3.30 \\
\bottomrule[1pt]
\end{tabular}
\label{tab:mos}
\end{table}
\begin{table*}[t]
\small
\centering
\caption{Objective performance of \ours{}$_{1:3}$ using the decoder in chunk-wise streaming mode. Once the generated tokens reach the sum of \textit{Chunk Size} and \textit{Right Context}, they are fed into the decoder, with \textit{Right Context} as lookahead.}
\renewcommand\tabcolsep{10pt}
\resizebox{0.85\linewidth}{!}{
\begin{tabular}{cccccc}
\toprule
\multirow{2}{*}{\textbf{Chunk Size}} & \multirow{2}{*}{\textbf{Right Context}} & \multicolumn{2}{c}{\textbf{Continuation}} & \multicolumn{2}{c}{\textbf{Cross-Sentence}} \\
\cmidrule(r){3-4} \cmidrule(r){5-6} & & WER-H$\downarrow$ & SIM-o$\uparrow$ & WER-H$\downarrow$ & SIM-o$\uparrow$ \\
\midrule
-  & -  & 3.60 & 0.757 & 4.53 & 0.653 \\
50 & 20 & 3.75 & 0.762 & 5.36 & 0.663 \\
25 & 10 & 3.74 & 0.753 & 5.50 & 0.651 \\
15 & 6  & 4.24 & 0.722 & 5.82 & 0.628 \\
\bottomrule
\end{tabular}
}
\label{tab:streaming}
\end{table*}

\section{Experiments}
\subsection{Experimental Setup}
\subsubsection{Dataset}
We conduct experiments on the LibriTTS~\cite{libritts} dataset, a multi-speaker English corpus with approximately 580 hours of speech from 2,306 speakers.
For text tokenization, we use 2,000-class BPE word pieces. Speech tokenization is carried out using the off-the-shelf S3Tokenizer model\footnote{\url{https://github.com/xingchensong/S3Tokenizer}} from CosyVoice~\cite{cosyvoice} at a token rate of 50Hz. Speech reconstruction is performed using the off-the-shelf OT-CFM model with the built-in vocoder, also from CosyVoice~\cite{cosyvoice}.

\subsubsection{Model}
We employ a decoder-only transformer architecture with 12 layers, 16 attention heads, 1024-dimensional embeddings, and 4096-dimensional feed-forward layers, with a total of 161.8M parameters.
All models are trained on 8 NVIDIA V100 32GB GPUs with a 160-second batch duration per GPU for 50 epochs.
We utilize the ScaledAdam~\cite{zipformer} optimizer and Eden~\cite{zipformer} scheduler, with a peak learning rate of 0.045.

\subsection{Evaluation}
\subsubsection{Evaluation Settings}
We use the LibriSpeech~\cite{librispeech} test-clean set for zero-shot TTS evaluation, ensuring no overlap in speakers with the training set. Following previous practice~\cite{valle}, the same test set is employed, which comprises audio segments ranging from 4 to 10 seconds, totaling 2.2 hours of data from 40 unique speakers and 1,234 samples.
We evaluate \ours{} under two inference tasks:
\begin{itemize}[leftmargin=*, itemsep=0pt, topsep=0pt]
\item \textit{Continuation}: Using the text transcription and the first 3 seconds of an utterance as a prompt, the model synthesizes the remainder of the speech;
\item \textit{Cross-Sentence}: Using a reference utterance and its transcription as the prompt, the model generates speech for the target text while preserving the characteristics of the speaker.
\end{itemize}

\subsubsection{Evaluation Metrics}
We employ the following objective metrics, including WER, SIM, and UTMOSv2, to assess the robustness, speaker similarity, overall perceived quality, and efficiency of the proposed method, respectively. For the continuation task, we evaluate the entire utterance rather than just the continuation segment for a more complete comparison.
\begin{itemize}[leftmargin=*, itemsep=0pt, topsep=0pt]
\item \textbf{WER-H} (Word Error Rate) is used to evaluate the robustness and intelligibility of synthesized speech. Neural TTS systems often encounter robustness issues.
To evaluate these, we perform speech recognition on the synthesized output using the HuBERT-Large~\cite{hubert} ASR model\footnote{\url{https://huggingface.co/facebook/hubert-large-ls960-ft}} and calculate the WER between the generated transcripts and the ground truth text.

\item \textbf{SIM-o} (Speaker Similarity) measures the similarity between the original prompt and synthesized speech. We use the state-of-the-art speaker verification model WavLM-TDNN\footnote{
\url{https://github.com/microsoft/UniSpeech/tree/main/downstreams/speaker_verification\#pre-trained-models}}~\cite{wavlm}. The similarity score predicted by WavLM-TDNN ranges from $[-1, 1]$, with a higher score indicating greater speaker similarity.

\item \textbf{UTMOSv2} measures the naturalness and overall quality of synthesized speech. We use the UTokyo-SaruLab Mean Opinion Score Prediction System v2 (UTMOSv2)~\cite{utmosv2}, a model-based, non-intrusive speech quality metric trained on human ratings. The predicted score ranges from 1 to 5, with higher scores denoting better perceptual quality. UTMOSv2 offers an efficient and reliable estimation of human judgment in speech synthesis evaluation.

\item \textbf{RTF} (Real-Time Factor) measures the time taken to synthesize one second of speech and reflects system efficiency, especially in real-time scenarios. We report RTF on an NVIDIA TESLA A100 80G GPU, calculated from the average inference time for generating 10 seconds of speech with a batch size of 1.
\end{itemize}

\subsubsection{Baseline Systems}
We evaluate our systems against several state-of-the-art (SOTA) zero-shot TTS systems.
For a fair comparison, we reproduce VALL-E using the same training data. In addition, we compare our systems with multiple SOTA systems, including MaskGCT, E2-TTS, and the original VALL-E trained on a large-scale dataset.
Note that our goal is not to pursue SOTA performance, but rather to comprehensively explore the proposed interleaved speech-text language modeling paradigm on a relatively small dataset.
More details about the baseline systems are provided in Appendix~\ref{sec:baselines}.

\subsection{Main Results}
Table~\ref{tab:main} presents comparisons between our proposed \ours{} and the baselines in terms of robustness, similarity, and efficiency on the LibriSpeech test-clean set. Table~\ref{tab:mos} reports the overall perceived quality of \ours{} compared to the baselines.

\subsubsection{Comparison with Baselines}
\ours{}$_{\infty:\infty}$ consistently outperforms two VALL-E variants across all evaluation metrics for both continuation and cross-sentence tasks, despite the reconstructed ground truth from S3Tokenizer being notably inferior in quality to that from EnCodec. \ours{} is based on 50Hz single-layer semantic speech tokens from S3Tokenizer, whereas VALL-E relies on 75Hz eight-layer acoustic speech tokens from EnCodec. This suggests that single-layer semantic representations are more amenable to effective modeling by language models.

Remarkably, despite being trained with much less data, \ours{}$_{\infty:\infty}$ outperforms the large-scale trained MaskGCT in both intelligibility and overall perceived quality, albeit with reduced similarity.
Furthermore, both our non-streaming \ours{}$_{\infty:\infty}$ and streaming \ours{}$_{1:3}$ achieve human-level overall perceived quality, outperforming the ground-truth recordings, MaskGCT, and E2 TTS on the cross-sentence task.
We also observe that the reconstructed ground-truth speech from the S3Tokenizer surpasses the original in overall perceived quality, which we attribute to flow matching in the speech detokenizer.

Compared to all baselines, \ours{} achieves the lowest RTF, which can be attributed to its compact design. Speech detokenization is performed once after all speech tokens are generated. The total number of generated tokens remains the same regardless of streaming mode or interleaving ratio. As a result, the RTF exhibits only minor variation and is reported as a single value.

\subsubsection{Comparison among Streaming Variants}
Among all streaming systems, \ours{}$_{1:3}$ achieves the best overall performance on both continuation and cross-sentence tasks.
Compared to its non-streaming counterpart \ours{}$_{\infty:\infty}$, \ours{}$_{1:3}$ exhibits a relatively small WER gap, specifically 6.94\% for continuation and 8.17\% for cross-sentence, and comparable similarity.
These results demonstrate that \ours{} effectively maintains performance for streaming without the need for complex engineering.

\subsubsection{Comparison under Chunk-wise Streaming}
Table \ref{tab:streaming} provides results for \ours{}$_{1:3}$ with the decoder in chunk-wise streaming mode. The model generates speech tokens concurrently with waveform synthesis, and the response latency is controlled by the chunk size and right context.
\begin{figure}[H]
\centering

\begin{subfigure}[t]{0.48\linewidth}
    \centering
    \includegraphics[width=\linewidth]{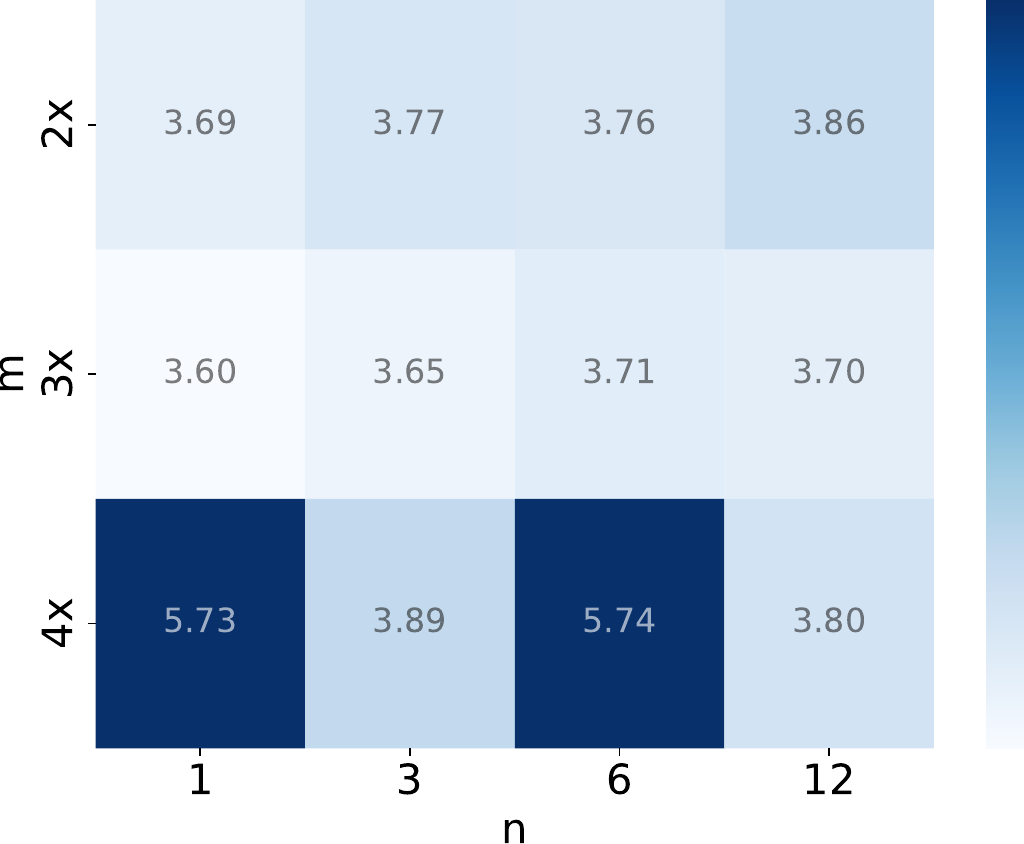}
    \subcaption{Continuation}
    \label{fig:heatmap_wer_conti}
\end{subfigure}
\hfill
\begin{subfigure}[t]{0.48\linewidth}
    \centering
    \includegraphics[width=\linewidth]{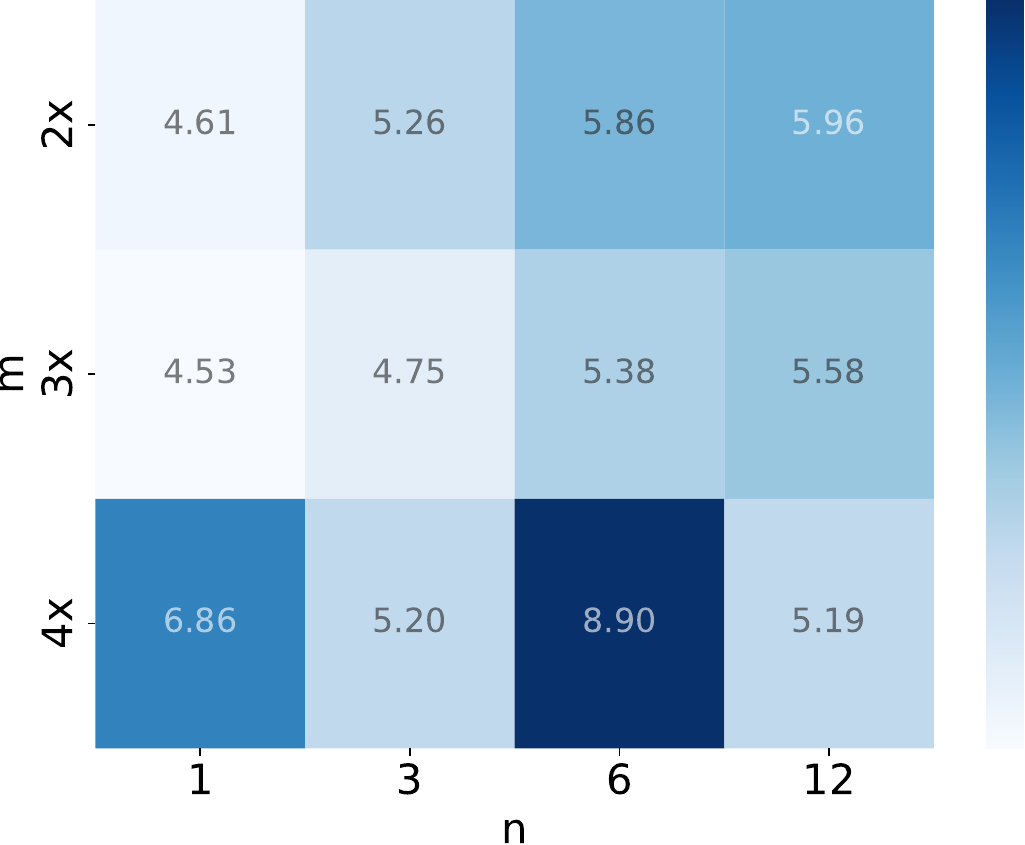}
    \subcaption{Cross-Sentence}
    \label{fig:heatmap_wer_cross}
\end{subfigure}

\caption{Heatmap of WER of continuation and cross-sentence tasks as the ratio of text chunk size $n$ to speech chunk size $m$ varies. The horizontal axis represents the text chunk size $n$, while the vertical axis represents the speech chunk size $m$. The color intensity reflects the magnitude of the WER values.}
\label{fig:heatmap_wer}
\end{figure}
\begin{figure*}[t]
\centering
\begin{subfigure}[t]{0.32\linewidth}
    \centering
    \includegraphics[width=\linewidth]{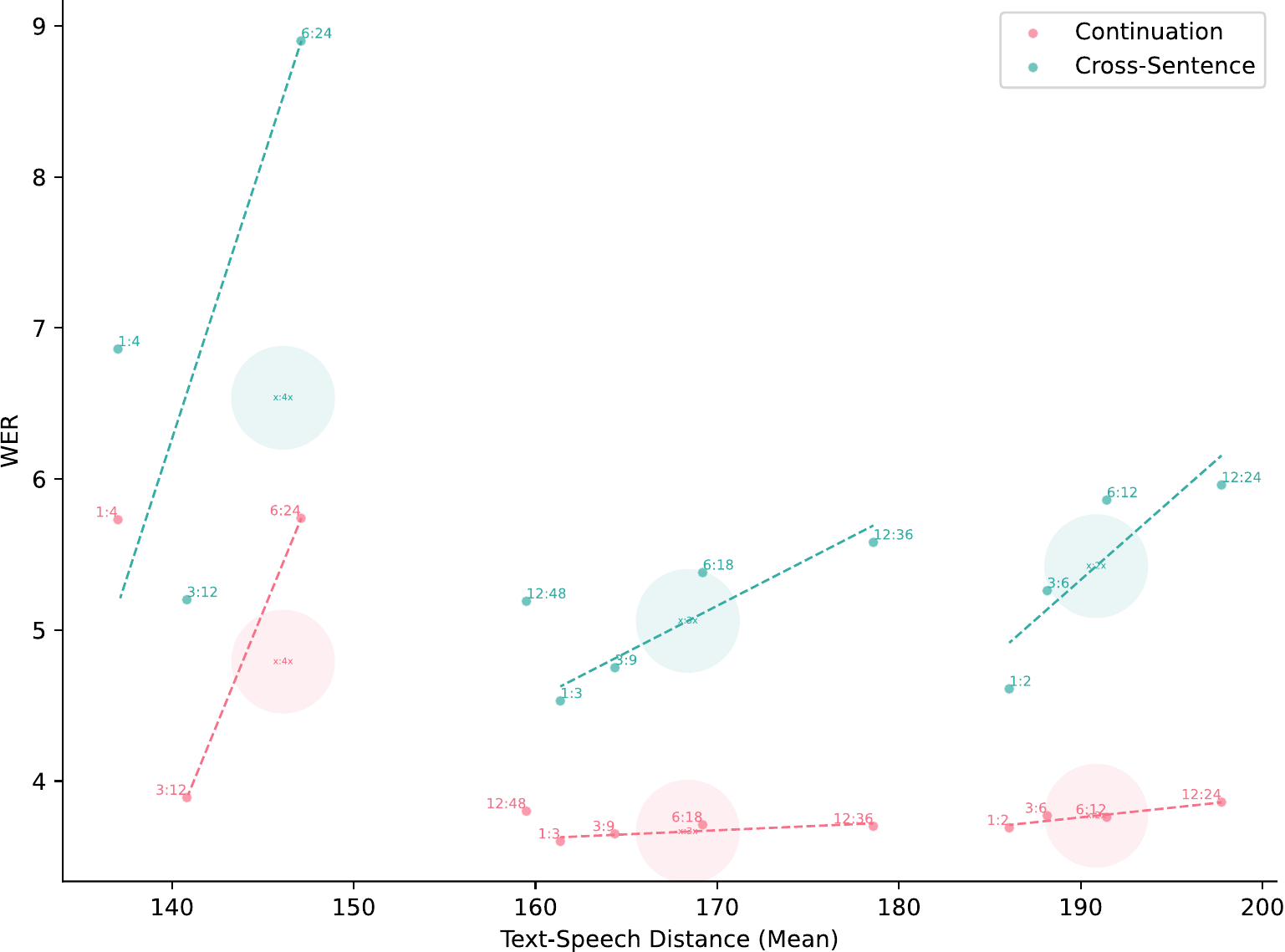}
    \subcaption{Average $\mu_{D}$}
    \label{fig:corr_dist_mean}
\end{subfigure}
\hfill
\begin{subfigure}[t]{0.32\linewidth}
    \centering
    \includegraphics[width=\linewidth]{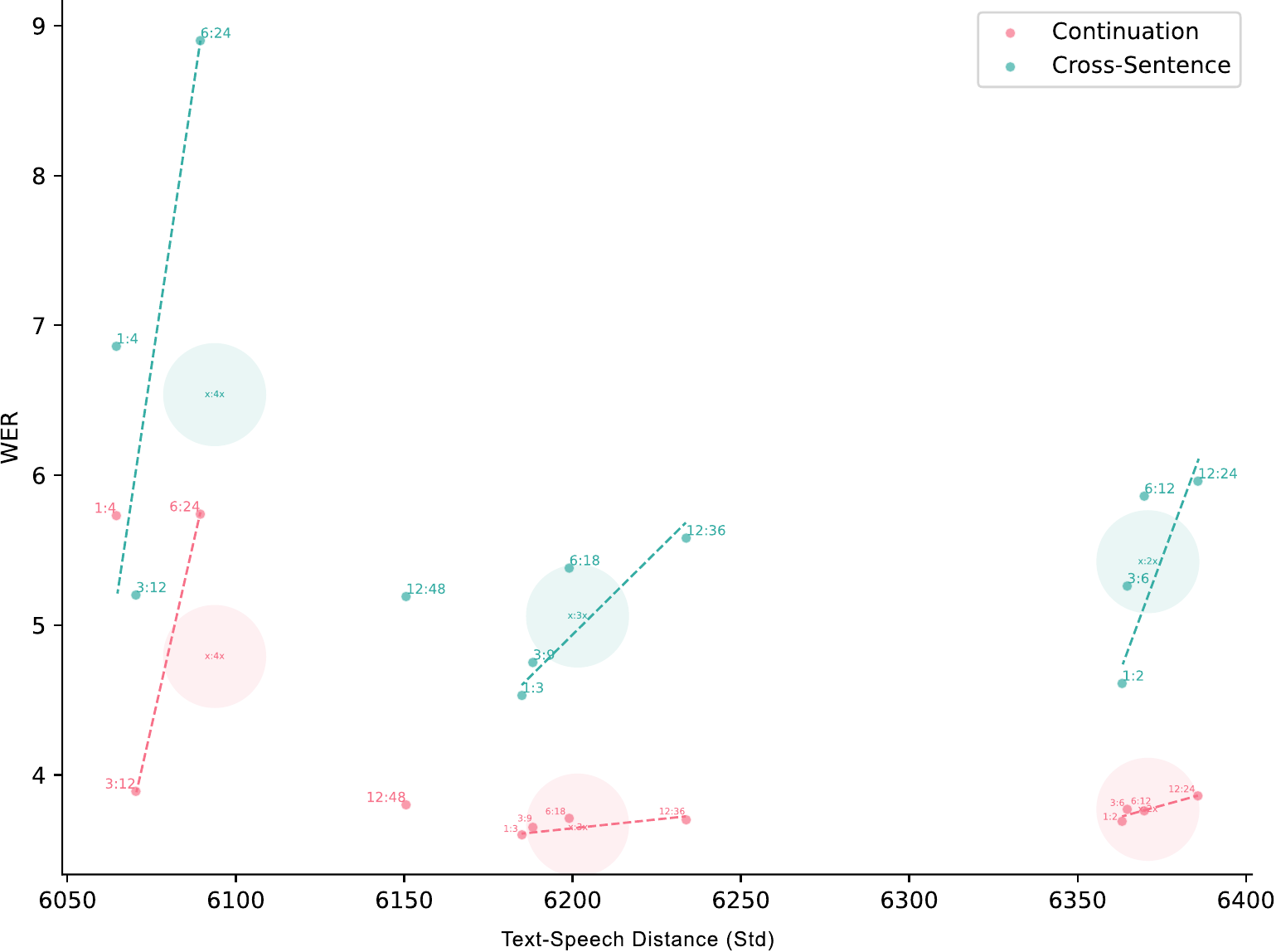}
    \subcaption{Average $\sigma_{D}$}
    \label{fig:corr_dist_std}
\end{subfigure}
\hfill
\begin{subfigure}[t]{0.32\linewidth}
\centering
\includegraphics[width=\linewidth]{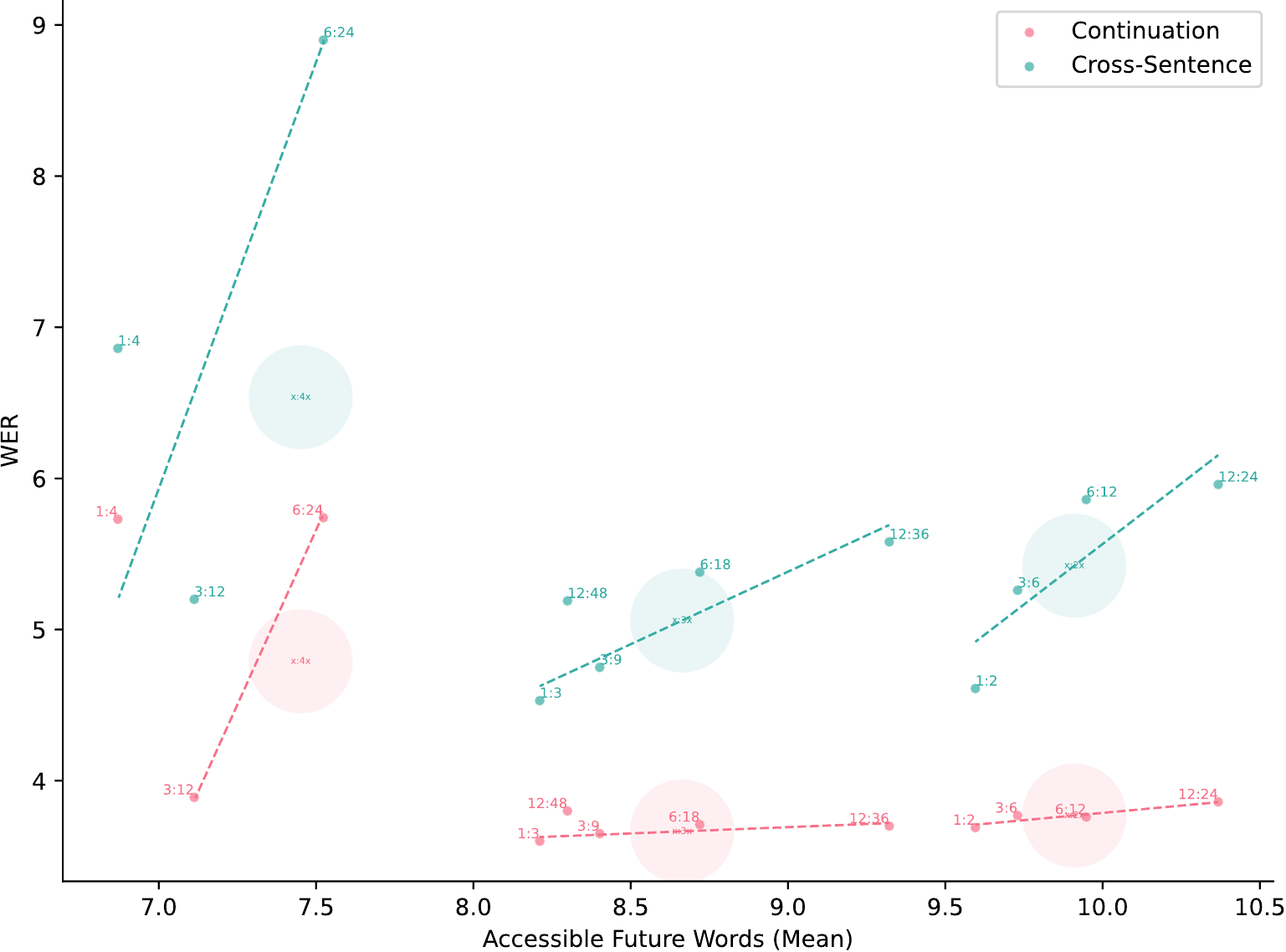}
\subcaption{Average $A$}
\label{fig:corr_future_mean}
\end{subfigure}

\caption{Correlation between the three statistical measures and the WERs of continuation and cross-sentence tasks. The WERs are grouped by the values of the ratio $n \colon m$, with the central points of each group represented by large circles ($x \colon 2x$, $x \colon 3x$, $x \colon 4x$). For each group, the four data points are fitted using Linear Regression with Random Sample Consensus (RANSAC), and the fitted lines are shown as dashed lines.}
\label{fig:correlation_analysis}
\end{figure*}

\section{Analyses}
\subsection{Impact of Ratio on Performance}
Fig.~\ref{fig:heatmap_wer} shows a heatmap of WERs for two tasks.
The horizontal axis represents text chunk size $n$, while the vertical axis represents speech chunk size $m$.
As $n$ increases, WER for both continuation and cross-sentence tasks generally increases, except for two noise outliers ($1\colon4$ and $12\colon48$), indicating that larger chunk sizes tend to have worse performance. Additionally, as the value of ratio $n:m$ increases, WER first decreases and then increases, reflecting the influence of multiple factors.

\subsection{Definitions of Position-Aware Measures}
To investigate the key factors involved in the interleaving design, including chunk-internal size and chunk-to-chunk ratio, we propose four sets of word-level, position-aware statistical measures.
Each training sample comprises up to 72 words. For each word $ j $ in sample $ i $, it can be encoded into multiple BPE tokens $ x_{ij}^{0}, x_{ij}^{1}, \dots, x_{ij}^{l_1} $, and corresponding speech tokens $ y_{ij}^{0}, y_{ij}^{1}, \dots, y_{ij}^{l_2} $ are obtained through word-level forced alignment. We define the distance between tokens $ x $ and $ y $ as $ d(x, y) $. The speech-text distance for word $ j $ in sample $i$, denoted $ D_{ij} $, is calculated as the average distance between each speech token and all corresponding BPE tokens: $ D_{ij} = \frac{1}{l_2} \sum_{k=1}^{l_2} \frac{1}{l_1} \sum_{r=1}^{l_1} d(x_{ij}^{r}, y_{ij}^{k}) $.
The mean and standard deviation of the speech-text distance for each word position $j$ across the entire training set are denoted as $ \mu_{D_j} $ and $ \sigma_{D_j}$, respectively.
Similarly, we define the average number of future words accessible by the speech tokens corresponding to each word position as $A_j$.
Additionally, we analyze the frequency with which the speech tokens corresponding to each word position precede the BPE tokens of the current word, denoted as $F_j$.
We perform statistical analyses on the training dataset using the above-mentioned measures.
Fig.~\ref{fig:visualization} visualizes $\mu_{D_j}$, $\sigma_{D_j}$, $A_j$, and $F_j$ for each word position $j$ across different ratio settings.

\subsection{Impact of Position-Aware Measures on Performance}
Fig.~\ref{fig:correlation_analysis} shows the correlation between average measures of all word positions and WERs for two tasks, leading to the following conclusions:
\begin{itemize}[leftmargin=*]
\item \textbf{Effect of $n \colon m$:} The ratio $n \colon m$ directly affects $\mu_{D_j}$, $\sigma_{D_j}$, $A_j$, and $F_j$. Specifically, when the value of ratio is fixed and $n$ (i.e., chunk-internal size) increases, both $\mu_{D_j}$ and $A_j$ increase, $\sigma_{D_j}$ slightly increases, and $F_j$ decreases. Conversely, when $n$ is fixed and the ratio (i.e., chunk-to-chunk ratio) increases, $\mu_{D_j}$ and $\sigma_{D_j}$ decrease, $A_j$ decreases, and $F_j$ increases.
\item \textbf{Effect of $\mu_{D_j}$, $\sigma_{D_j}$, $A_j$:} When $\mu_{D_j}$ increases, $\sigma_{D_j}$ and $A_j$ also increases. The WER for both continuation and cross-sentence tasks first decreases and then increases. This reflects a trade-off, where shorter distances impose stronger constraints on speech synthesis, limiting contextual information as fewer upcoming text tokens are accessible to the current speech token while increasing the modeling difficulty for the LM.
\item \textbf{Effect of $F_j$:} The frequency of speech tokens preceding text tokens occurs mainly at the start of the interleaved sequence when $n$ is small and the ratio is large. This increases training difficulty, as the speech tokens lack text context, but typically do not affect inference with the speech prompt, except for the $1\colon4$ ratio, which exhibits abnormally high WERs.
\item \textbf{Outlier analysis:} \ours{}$_{12:48}$ exhibits abnormally low WERs, as around 40\% of test samples in the continuation task contain no more than 24 text tokens, resembling non-streaming behavior.
\end{itemize}
\section{Conclusion}
This paper introduces \ours{} for zero-shot streaming TTS, which is directly trained on interleaved text and speech tokens at a fixed ratio. Experiments on LibriTTS demonstrate that \ours{} with a $1\colon3$ ratio significantly outperforms other streaming systems, achieving acceptably worse intelligibility compared to non-streaming counterpart while maintaining comparable speaker similarity and overall perceived quality. Furthermore, our analysis provides several insights into how the ratio impacts performance, revealing the trade-offs between enforcing textual constraints and leveraging contextual information in speech synthesis. We hope that the language modeling paradigm of \ours{} and the insights gained from our analysis will contribute to advancing the field of voice interaction.


\section*{Limitations}
Despite the promising performance and compact topology, we acknowledge several limitations. This work initially employed a non-streaming decoder and simulated streaming inference by chunking speech tokens, due to the unavailability of an off-the-shelf streaming decoder at the time. This led to first-packet latency constrained by the chunk size and degraded speech quality. We anticipate that performance will improve with an advanced streaming decoder.

\bibliography{custom}

\appendix
\section{Details of Baselines}
\label{sec:baselines}

\begin{itemize}[leftmargin=*, itemsep=0pt, topsep=0pt]
\item \textbf{VALL-E}~\cite{valle}: A two-stage TTS system that includes both autoregressive (AR) and non-autoregressive (NAR) models to generate RVQ tokens at 75Hz, based on EnCodec~\cite{encodec}. We consider two VALL-E variants: (1) the version trained on the Librilight corpus~\cite{librilight}, for which we use the performance results from the original paper~\cite{valle} and RTF reported in~\cite{melle} using the official checkpoint; and (2) a reproduction trained on LibriTTS~\cite{libritts}, using the publicly available codebase\footnote{\url{https://github.com/lifeiteng/vall-e}}.

\item \textbf{E2 TTS}~\cite{e2tts}: A fully NAR system based on flow-matching, comprising 333M parameters. We use the publicly available pre-trained checkpoint\footnote{\url{https://huggingface.co/SWivid/E2-TTS}}, trained on 100K hours of in-the-wild Chinese and English data from the Emilia corpus~\cite{emilia}.

\item \textbf{MaskGCT}~\cite{maskgct}: A fully NAR two-stage TTS system based on masked language modeling, comprising a 695M text-to-semantic model and a 353M semantic-to-acoustic model. We use the official pre-trained checkpoint\footnote{\url{https://huggingface.co/amphion/MaskGCT}}, trained on 100K hours of in-the-wild Chinese and English data from Emilia corpus~\cite{emilia}.
\end{itemize}

\section{Visualization of Statistical Measures}
\label{sec:visualization}
\begin{figure*}[t]
\centering

\begin{subfigure}[t]{0.2\linewidth}
    \centering
    \includegraphics[height=0.85\textheight]{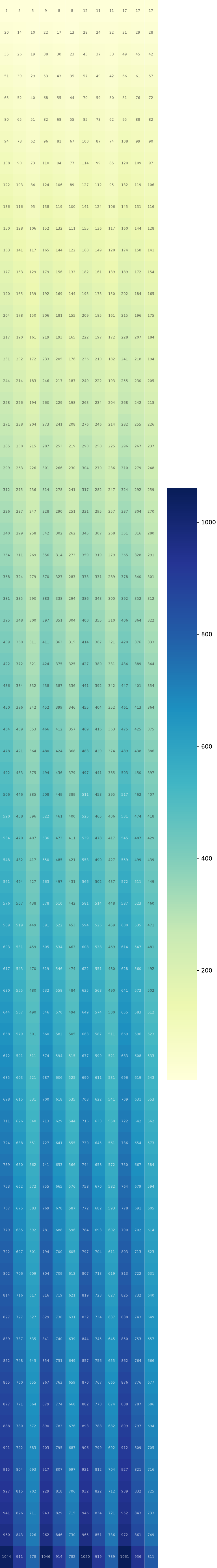}
    \subcaption{Text-Speech Distance (Mean)}
    \label{fig:dist_mean}
\end{subfigure}
\hfill
\begin{subfigure}[t]{0.2\linewidth}
    \centering
    \includegraphics[height=0.85\textheight]{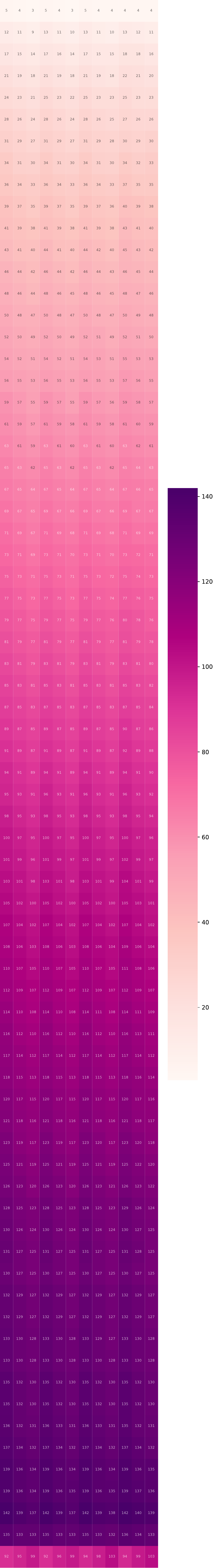}
    \subcaption{Text-Speech Distance (Std)}
    \label{fig:dist_std}
\end{subfigure}
\hfill
\begin{subfigure}[t]{0.2\linewidth}
    \centering
    \includegraphics[height=0.85\textheight]{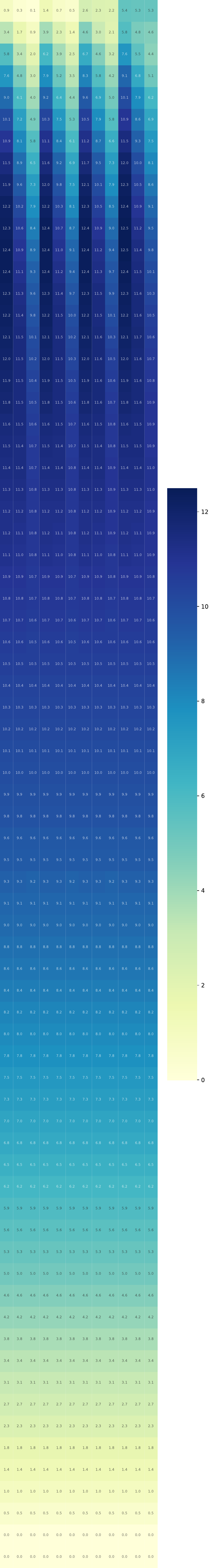}
    \subcaption{Accessible Future Words}
    \label{fig:future_mean}
\end{subfigure}
\hfill
\begin{subfigure}[t]{0.2\linewidth}
    \centering
    \includegraphics[height=0.85\textheight]{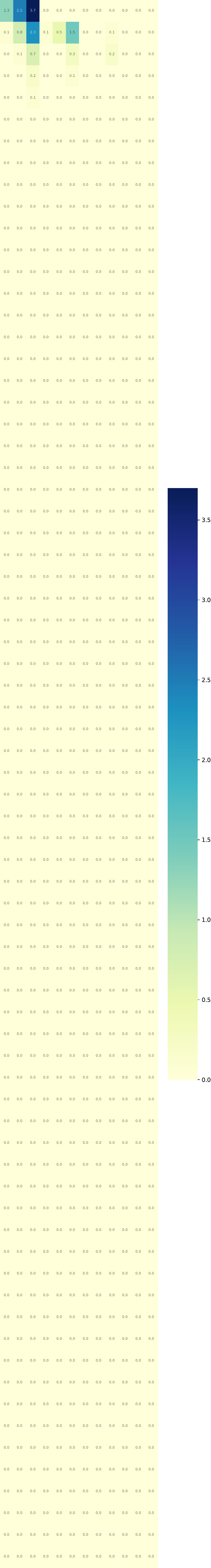}
    \subcaption{Unbound Speech Tokens}
    \label{fig:precedence_mean}
\end{subfigure}
    
\caption{Visualization of four statistical measures. From left to right in each plot, the ratios are $1\colon2$, $1\colon3$, $1\colon4$, $3\colon6$, $3\colon9$, $3\colon12$, $6\colon12$, $6\colon18$, $6\colon24$, $12\colon24$, $12\colon36$, and $12\colon48$. From top to bottom, the plots correspond to the first through the 72nd word. The color intensity reflects the magnitude of the values.}
\label{fig:visualization}
\end{figure*}

Fig.~\ref{fig:visualization} presents heatmaps illustrating the values of four statistical metrics across different positions (from top to bottom) and varying ratios of text chunk size to speech chunk size (from left to right). Darker colors indicate higher values.

\end{document}